\documentclass[aps,prd,superscriptaddress,twoside,twocolumn,nofootinbib,showpacs,floatfix]{revtex4-1}
\usepackage{amsmath,amssymb}
\usepackage{graphicx,bm,braket}
\usepackage{subfigure}
\allowdisplaybreaks
\pdfpagewidth=\paperwidth
\pdfpageheight=\paperheight
\allowdisplaybreaks
\usepackage[normalem]{ulem} 
\usepackage[usenames]{color}
\renewcommand\sout{\bgroup \color{red} \ULdepth=-.5ex \ULset}

\usepackage{ulem}

\begin{document}

\title{\boldmath Photoproduction reaction $\gamma n \to K^{\ast 0}\Lambda$ in an effective Lagrangian approach}

\author{Neng-Chang Wei}
\affiliation{School of Nuclear Science and Technology, University of Chinese Academy of Sciences, Beijing 101408, China}

\author{Yi-Ming Zhu}
\affiliation{School of Physical Sciences, University of Chinese Academy of Sciences, Beijing 101408, China}

\author{Fei Huang}
\email[Corresponding author. Email: ]{huangfei@ucas.ac.cn}
\affiliation{School of Nuclear Science and Technology, University of Chinese Academy of Sciences, Beijing 101408, China}

\date{\today}

\begin{abstract}
In our previous work [Phys. Rev. C 101, 014003 (2020)], the photoproduction reaction $\gamma p \to K^{\ast +} \Lambda$ has been investigated within an effective Lagrangian approach. There, the reaction amplitudes were constructed by including the $t$-channel $K$, $K^\ast$, and $\kappa$ exchanges, the $u$-channel $\Lambda$, $\Sigma$, and $\Sigma^\ast$ exchanges, the $s$-channel $N$, $N(2000)5/2^+$, and $N(2060)5/2^-$ exchanges, and the interaction current. It has been shown that the data on both the differential cross sections and the spin density matrix elements were simultaneously and satisfactorily described. In this paper, we study the photoproduction reaction $\gamma n \to K^{\ast 0} \Lambda$ based on the same reaction mechanism as that of $\gamma p \to K^{\ast +} \Lambda$ with the purpose of getting a unified description of the data for both $\gamma p \to K^{\ast +} \Lambda$ and $\gamma n \to K^{\ast 0} \Lambda$ within a same model. All hadronic coupling constants, form factor cutoffs, and the resonance masses and widths in the present calculations remain the same as in our previous work for $\gamma p \to K^{\ast +} \Lambda$. The available differential cross-section data for $\gamma n \to K^{\ast 0} \Lambda$ are well reproduced. Further analysis shows that the cross sections of $\gamma n \to K^{\ast 0} \Lambda$ are dominated by the contributions of the $t$-channel $K$ exchange, while the $s$-channel $N(2000)5/2^+$ and $N(2060)5/2^-$ exchanges provide considerable contributions as well.
\end{abstract}


\maketitle


\section{Introduction}\label{sec:Intro}

The study of the nucleon resonances ($N^\ast$'s) has always been a topic of great interest in hadron physics, since a deeper understanding of the nucleon resonances is essential to get insight into the nonperturbative regime of quantum chromodynamics (QCD). Our current knowledge of most of the $N^\ast$'s is mainly coming from the production reactions of $\pi N$, $\eta N$, $K \Lambda$, and $K\Sigma$ channels. In recent years, the photoproductions of vector mesons, $\eta'$ meson, and $KY^\ast$ $(Y=\Lambda,\Sigma)$ have also been extensively investigated both experimentally and theoretically towards getting a better understanding of $N^\ast$'s \cite{Wang:2017tpe,Kim:2014hha,Wang:2018vlv,Wang:2019mid,Wei:2020,Wei:2019imo,Zhang2021etap,Anisovich:2018yoo,Tiator:2018heh,Moriya2013,Wang2020,Wei2021L1520,WXLGO2017,KimNam2017,Zhang2021L1405}. 

In the present work, we are interested in the photoproduction reaction $\gamma n \to K^{\ast 0} \Lambda$. The threshold of the $K^\ast\Lambda$ photoproduction is around center-of-mass energy $W\approx 2.0$ GeV, and thus makes this reaction more suitable to study the less-explored high-mass resonances. What's more, the isospin $I=1/2$ for the final states $K^\ast\Lambda$ forbids the $s$-channel $I=3/2$ $\Delta$ resonance exchanges to contribute, thus provides facilities for the extraction of the information on the $I=1/2$ nucleon resonances.

Experimentally, unlike the reaction $\gamma p \to K^{\ast +} \Lambda$ for which we have high-precision differential cross-section data \cite{Tang:2013} and the data on spin density matrix elements \cite{Anisovich:2017rpe}, the only data available for $\gamma n \to K^{\ast 0} \Lambda$ are differential cross sections at three photon energies in the range $1.9 < E_\gamma < 2.5$ GeV from the CLAS Collaboration \cite{Mattione:2011pya}. 

Theoretically, the CLAS differential cross-section data for $\gamma n \to K^{\ast 0} \Lambda$ \cite{Mattione:2011pya} have been so far analyzed in two publications \cite{Wan:2015gsl,Yu:2016spg}. In Ref.~\cite{Wan:2015gsl}, the reaction $\gamma n \to K^{\ast 0} \Lambda$ was studied by use of an effective Lagrangian approach, where no resonance exchanges were considered and the data were described by adjusting the cutoff parameter of the $t$-channel form factor. It was reported that the differential cross sections of $\gamma n \to K^{\ast 0} \Lambda$ \cite{Mattione:2011pya} were overwhelmingly dominated by the $t$-channel $K$ exchange, while the contributions from all other terms were totally negligible. Although the cross-section data for $\gamma n \to K^{\ast 0} \Lambda$ have been qualitatively described in Ref.~\cite{Wan:2015gsl}, it is not clear whether the employed interactions, especially the overwhelmingly dominated $t$-channel $K$ exchange, works simultaneously for the reaction $\gamma p \to K^{\ast +} \Lambda$, of which not only the high-precision differential cross-section data but also the data on spin density matrix elements are available \cite{Tang:2013,Anisovich:2017rpe}. In Ref.~\cite{Yu:2016spg}, the differential cross-section data for both $\gamma p \to K^{\ast +} \Lambda$ and $\gamma n \to K^{\ast 0} \Lambda$ \cite{Tang:2013,Mattione:2011pya} were simultaneously analyzed by considering the $t$-channel $K$, $K^\ast$, and $\kappa$ exchanges within a Regge model. It was stated that the $t$-channel $K$ exchange dominated the differential cross sections in both reactions. Nevertheless, due to the lack of contributions from the $s$-channel nucleon resonance exchanges, the angular distribution data for both the $\gamma p \to K^{\ast +} \Lambda$ and $\gamma n \to K^{\ast 0} \Lambda$ reactions were only qualitatively described in Ref.~\cite{Yu:2016spg}. One also notices that the data on spin density matrix elements for $\gamma p \to K^{\ast +} \Lambda$ \cite{Anisovich:2017rpe} had not been considered in the analysis of Ref.~\cite{Yu:2016spg}.

As the hadronic vertices and propagators are exactly the same in both $\gamma p \to K^{\ast +} \Lambda$ and $\gamma n \to K^{\ast 0} \Lambda$ except for some possible isospin factors, and most of the electromagnetic coupling constants can be determined by the radiative decays of the corresponding hadrons, a combined analysis of the available differential cross-section data for both $\gamma p \to K^{\ast +} \Lambda$ and $\gamma n \to K^{\ast 0} \Lambda$ and the data on spin density matrix elements for $\gamma p \to K^{\ast +} \Lambda$ is of great interest and significance. It provides more constraints on the theoretical model, and moreover, it makes the analysis of the data for $\gamma n \to K^{\ast 0} \Lambda$  more reliable in the situation that so far only the differential cross-section data at three energy points are available for this reaction.

In Ref.~\cite{Wei:2020}, we have studied the photoproduction reaction $\gamma p \to K^{\ast +} \Lambda$ in an effective Lagrangian approach. There, apart from the $t$-channel $K$, $K^\ast$, and $\kappa$ exchanges, the $u$-channel $\Lambda$, $\Sigma$, and $\Sigma^\ast$ exchanges, the $s$-channel $N$ exchange, and the interaction current, we considered in the $s$ channel as few as possible nucleon resonance exchanges in constructing the reaction amplitudes to describe the data. The gauge invariance of the photoproduction amplitudes was fully implemented. It was found that by introducing the $N(2060)5/2^-$ and $N(2000)5/2^+$ resonance exchanges, the available data on both differential cross sections and spin density matrix elements for $\gamma p \to K^{\ast +} \Lambda$ \cite{Tang:2013,Anisovich:2017rpe} can be reasonably reproduced. The $t$-channel $K$ exchange and $s$-channel $N(2060)5/2^-$ and $N(2000)5/2^+$ exchanges were found to provide dominant contributions for this reaction.

In this work, we investigate the photoproduction reaction $\gamma n \to K^{\ast 0} \Lambda$ based on the same reaction mechanism as that of $\gamma p \to K^{\ast +} \Lambda$ in our previous work \cite{Wei:2020}. The purpose is to get a unified description of all the available differential cross-section data for both $\gamma p \to K^{\ast +} \Lambda$ and $\gamma n \to K^{\ast 0} \Lambda$ and the data on spin density matrix elements for $\gamma p \to K^{\ast +} \Lambda$ by use of the same reaction model. 

The paper is organized as follows. In the next section, we briefly introduce the framework of our theoretical model. The numerical results are shown and discussed in Sec.~\ref{Sec:results}. The summary and conclusions are given in Sec.~\ref{Sec:summary}.

\section{Formalism}  \label{Sec:formalism}

\begin{figure}
\centering
\subfigure[~$s$ channel]{\includegraphics[width=0.45\linewidth]{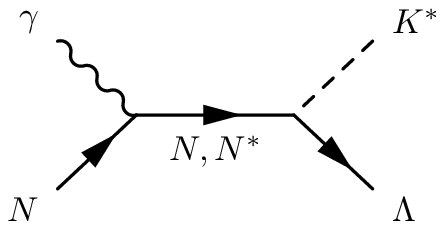}}
\hspace{0.01\linewidth}
\subfigure[~$t$ channel]{\includegraphics[width=0.45\linewidth]{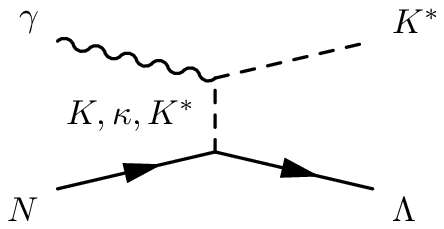}}
\vfill
\subfigure[~$u$ channel]{\includegraphics[width=0.45\linewidth]{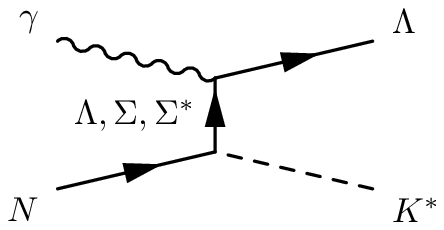}}
\hspace{0.01\linewidth}
\subfigure[~Interaction current]{\includegraphics[width=0.45\linewidth]{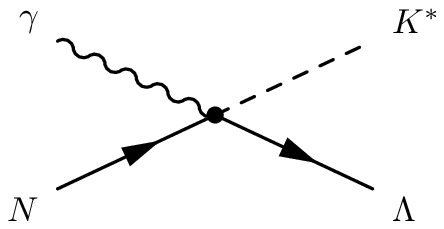}}
\caption{Generic structures of the photoproduction amplitudes for $\gamma p \to K^{\ast +} \Lambda$ and $\gamma n \to K^{\ast 0} \Lambda$. Time proceeds from left to right. For $\gamma n \to K^{\ast 0} \Lambda$, the interaction current and the $t$-channel $K^\ast$ exchange vanish.}
\label{FIG:feymans}
\end{figure}

The generic structures of the photoproduction amplitudes for $\gamma p \to K^{\ast +} \Lambda$ and $\gamma n \to K^{\ast 0} \Lambda$ in our effective Lagrangian approach are diagrammatically depicted in Fig.~\ref{FIG:feymans} \cite{Wang:2017tpe,Wang:2019mid,Wei:2020}. Specifically, we consider the $t$-channel $K$, $K^\ast$, and $\kappa$ exchanges, the $u$-channel $\Lambda$, $\Sigma$, and $\Sigma^\ast$ exchanges, the $s$-channel $N$ and $N^\ast$ exchanges, and the interaction current in constructing the reaction amplitudes for the $\gamma p \to K^{\ast +} \Lambda$ and $\gamma n \to K^{\ast 0} \Lambda$ reactions. The $s$-, $t$-, and $u$-channel amplitudes can be obtained straightforwardly by calculating the corresponding Feynman diagrams. The interacting current consists of the traditional Kroll-Ruderman term and an auxiliary current, with the later being constructed in such a way that the full photoproduction amplitudes satisfy the generalized Ward-Takahashi identity and thus is fully gauge invariant \cite{Wang:2017tpe}. For the photoproduction reaction $\gamma n \to K^{\ast 0} \Lambda$, the interaction current and the $t$-channel $K^\ast$ exchange vanish automatically due to the neutral charges of $K^{\ast 0}$ and $n$. The reaction amplitudes are apparently transverse in this case, and thus the requirement of gauge invariance on the production amplitudes is already fulfilled.

The effective Lagrangians, the resonance propagators, and the phenomenological form factors for the photoproduction reaction $\gamma N \to K^{\ast} \Lambda$ have been explicitly given in Ref.~\cite{Wang:2017tpe}, where we have analyzed the differential cross-section data for $\gamma p \to K^{\ast +} \Lambda$ in an effective Lagrangian approach, and we do not repeat them here for the sake of brevity. For the $\gamma n \to K^{\ast 0} \Lambda$ reaction which we study in the present work, the hadronic coupling constants, the form factor cutoffs, and the resonance masses and widths remain the same as that of $\gamma p \to K^{\ast +} \Lambda$. We quote them from our recent work of Ref.~\cite{Wei:2020}, which was an update of the work of Ref.~\cite{Wang:2017tpe}, where we have simultaneously analyzed the differential cross-section data and the data on spin density matrix elements for $\gamma p \to K^{\ast +} \Lambda$. 

The electromagnetic coupling constants in the reaction $\gamma n \to K^{\ast 0} \Lambda$ are, in principle, different from those in $\gamma p \to K^{\ast +} \Lambda$ when the $N$ or $K^{\ast}$ are involved in the electromagnetic vertices. For the $t$-channel $\kappa$ exchange, the coupling constant $g_{\gamma \kappa^0 K^{\ast 0}}=-0.428$ is taken from Refs.~\cite{Kim:2014hha,Wang:2018vlv}, determined by a vector-meson dominance model proposed by Black {\it et al.} \cite{Black:2002ek}. For the $t$-channel $K$ exchange, the coupling constant $g_{\gamma K^0 K^{\ast 0}}=-0.631$ is determined by the decay width of $K^{\ast 0} \to K^0 \gamma$ given by the Review of Particle Physics (RPP) \cite{Zyla:2020zbs} with the sign inferred from $g_{\gamma \pi \rho}$ \cite{Garcilazo:1993av} via the flavor SU(3) symmetry considerations in conjunction with the vector-meson dominance assumption. In the $s$ channel, apart from the $N$ exchange, it was found in our previous work \cite{Wei:2020} that the $N(2000)5/2^+$ and $N(2060)5/2^-$ exchanges are necessarily needed to describe the available differential cross-section data and the data on spin density matrix elements for $\gamma p \to K^{\ast +} \Lambda$. Here the same resonances are considered in the reaction $\gamma n \to K^{\ast 0} \Lambda$, with their electromagnetic coupling constants being treated as fit parameters since there is no experimental information for the helicity amplitudes of the $N(2000)5/2^+ \to n\gamma$ and $N(2060)5/2^- \to n\gamma$ decays. 

Actually, in tree-level calculations as performed in the present work and in Refs.~\cite{Wang:2017tpe,Wang:2019mid,Wei:2020}, only the products of the electromagnetic and hadronic coupling constants $g_{RN \gamma}^{(1,2)} g_{R\Lambda K^\ast}^{(1,2,3)}$ of each resonance can be uniquely determined. As the ratios of $g_{R\Lambda K^\ast}^{(2)}/g_{R\Lambda K^\ast}^{(1)}$ and $g_{R\Lambda K^\ast}^{(3)}/g_{R\Lambda K^\ast}^{(1)}$ for both $N(2060)5/2^-$ and $N(2000)5/2^+$ exchanges have already been determined in our previous work for the study of $\gamma p \to K^{\ast +} \Lambda$ \cite{Wei:2020}, here for the $\gamma n \to K^{\ast 0} \Lambda$ reaction, only the products $g_{RN \gamma}^{(1)} g_{R\Lambda K^\ast}^{(1)}$ and $g_{RN \gamma}^{(2)} g_{R\Lambda K^\ast}^{(1)}$ are left as adjustable parameters, which will be determined by a fit to the available data for this reaction.

\section{Results and discussion}   \label{Sec:results}

\begin{table}[tb]
\caption{\label{para:fit} Fitted values of adjustable parameters in the $\gamma n \to K^{\ast 0} \Lambda$ reaction.}
\begin{tabular*}{\columnwidth}{@{\extracolsep\fill}lcc}
\hline\hline
&  $N(2000)5/2^+$    & $N(2060)5/2^-$ \\
$g_{RN \gamma}^{(1)} g_{R\Lambda K^\ast}^{(1)}$ &   $-54.26\pm0.74$  &  $-8.13\pm0.51$\\
$g_{RN \gamma}^{(2)} g_{R\Lambda K^\ast}^{(1)}$ &   $-27.98\pm0.72$  &  $8.46\pm0.53$ \\
\hline\hline
\end{tabular*}
\end{table}

\begin{figure*}[tbp]
\includegraphics[width=0.8\textwidth]{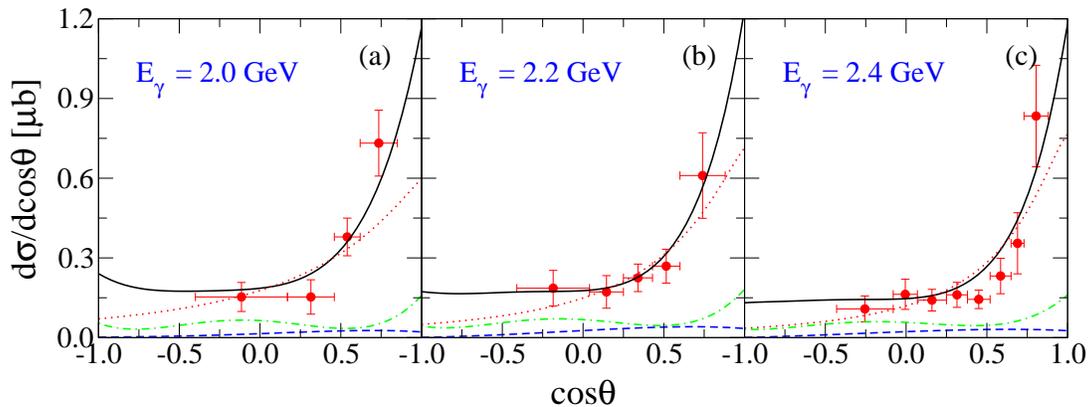}
\caption{Differential cross sections for $\gamma n \to K^{\ast 0} \Lambda$ as a function of $\cos\theta$ with $\theta$ being the scattering angle in center-of-mass frame. The black solid lines represent the results from the full calculation. The red dotted, green dash-dotted, and blue dashed lines represent the individual contributions from the $t$-channel $K$ exchange, the $s$-channel $N(2060)5/2^-$ exchange, and the $s$-channel $N(2000)5/2^+$ exchange, respectively. Data are taken from the CLAS Collaboration \cite{Mattione:2011pya}.}
\label{fig:dsdc}
\end{figure*}

\begin{figure*}[tbp]
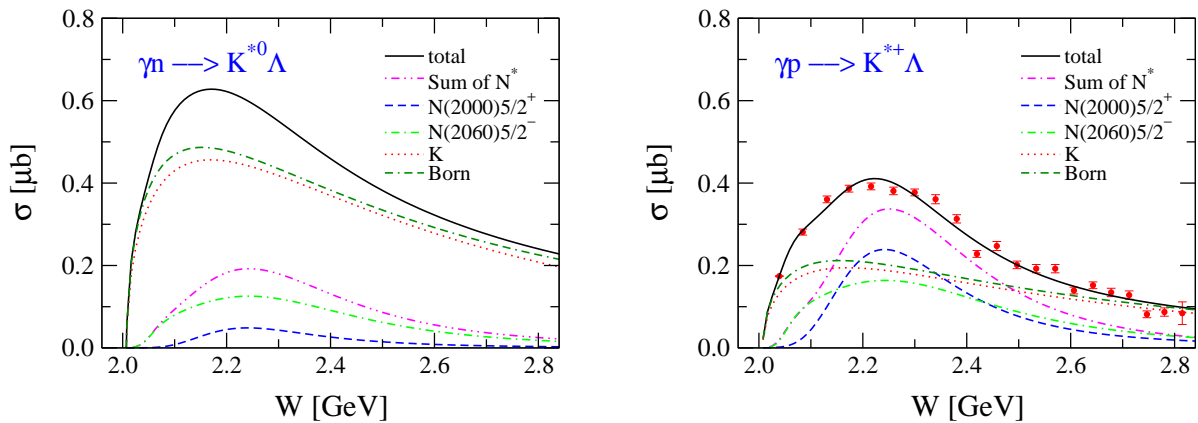

\vglue 0.25cm
\includegraphics[width=0.85\columnwidth]{total1.eps} 
\hglue 1.0cm
\includegraphics[width=0.85\columnwidth]{total.eps}
\caption{Predicted total cross sections with dominant individual contributions for $\gamma n \to K^{\ast 0} \Lambda$ (left) and $\gamma p \to K^{\ast +} \Lambda$ (right). Data for $\gamma p \to K^{\ast +} \Lambda$ are taken from the CLAS Collaboration \cite{Tang:2013} but not included in the fit.}
\label{fig:total}
\end{figure*}

The hadronic vertices and propagators in both the $\gamma p \to K^{\ast +} \Lambda$ and $\gamma n \to K^{\ast 0} \Lambda$ reactions are the same except for some possible isospin factors, and the electromagnetic couplings in these two reactions, in principle, can be determined by the radiative decays of the corresponding hadrons. Therefore, a unified description of all the available data for both $\gamma p \to K^{\ast +} \Lambda$ and $\gamma n \to K^{\ast 0} \Lambda$ is necessary and interesting. It puts more constraints on the theoretical model and results in a more reliable understanding of the reaction mechanisms for both $\gamma p \to K^{\ast +} \Lambda$ and $\gamma n \to K^{\ast 0} \Lambda$.

In our previous work \cite{Wei:2020}, we have investigated the photoproduction reaction $\gamma p \to K^{\ast +} \Lambda$ in an effective Lagrangian approach. By considering the $t$-channel $K$, $K^\ast$, and $\kappa$ exchanges, the $u$-channel $\Lambda$, $\Sigma$, and $\Sigma^\ast$ exchanges, the $s$-channel $N$, $N(2060)5/2^-$, and $N(2000)5/2^+$ exchanges, and the interaction current in constructing the reaction amplitudes, we have satisfactorily reproduced the high-precision differential cross section data and the data on spin density matrix elements for $\gamma p \to K^{\ast +} \Lambda$. It was found that the $t$-channel $K$ exchange and $s$-channel $N(2060)5/2^-$ and $N(2000)5/2^+$ exchanges provide dominant contributions for this reaction.

In this paper, we study the photoproduction reaction $\gamma n \to K^{\ast 0} \Lambda$ based on the same reaction mechanism as that of $\gamma p \to K^{\ast +} \Lambda$ in our previous work \cite{Wei:2020}. The interaction current and the $t$-channel $K^\ast$ exchange vanish automatically due to the neutral charges of $K^{\ast 0}$ and $n$. The hadronic coupling constants, the form factor cutoffs, and the resonance masses and widths for $\gamma n \to K^{\ast 0} \Lambda$ are the same as those for $\gamma p \to K^{\ast +} \Lambda$. The only adjustable parameters enter in the calculation of the amplitudes for $\gamma n \to K^{\ast 0} \Lambda$ are the products of the resonance hadronic and electromagnetic coupling constants, $g_{RN \gamma}^{(1)} g_{R\Lambda K^\ast}^{(1)}$ and $g_{RN \gamma}^{(2)} g_{R\Lambda K^\ast}^{(1)}$, which are determined by a fit to the available differential cross-section data for $\gamma n \to K^{\ast 0} \Lambda$. The fitted values for the $N(2060)5/2^-$ and $N(2000)5/2^+$ resonances are listed in Table~\ref{para:fit}. The corresponding results of the differential cross sections for $\gamma n \to K^{\ast 0} \Lambda$ are plotted in Fig.~\ref{fig:dsdc}. 

In Fig.~\ref{fig:dsdc}, the black solid lines represent the results from the full calculation. The red dotted, green dash-dotted, and blue dashed lines stand for the individual contributions from the $t$-channel $K$ exchange, the $s$-channel $N(2060)5/2^-$ exchange, and the $s$-channel $N(2000)5/2^+$ exchange, respectively. The contributions from the other individual terms are too small to be clearly seen with the scale used, and thus they are not plotted in this figure. One sees from Fig.~\ref{fig:dsdc} that our calculated differential cross sections for $\gamma n \to K^{\ast 0} \Lambda$ agree quite well with the experimental data. The $t$-channel $K$ exchange dominates the angular distributions in all those three energy points considered. In particular, it is responsible for the peaks of the differential cross sections at forward angles. The $s$-channel $N(2060)5/2^-$ exchange provides significant contributions, and considerable contributions from the $s$-channel $N(2000)5/2^+$ exchange are also observed.

In Ref.~\cite{Wan:2015gsl}, the differential cross sections of $\gamma n \to K^{\ast 0} \Lambda$ are almost fully described by the $t$-channel $K$ exchange. Here in the present work, we have much smaller contributions from the $t$-channel $K$ exchange. The difference is coming from the $t$-channel form factors. In Ref.~\cite{Wan:2015gsl} a monopole form factor is used, and the cutoff parameter is fixed by the $\gamma n \to K^{\ast 0} \Lambda$ differential cross section data, which gives $\Lambda_K=1050$ MeV. In the present work, a dipole form factor with the cutoff parameter $\Lambda_K=1009$ MeV is employed, which has been determined by the high-precision differential cross-section data and the data on spin density matrix elements for $\gamma p \to K^{\ast +} \Lambda$ in our previous work \cite{Wei:2020}. 

In Ref.~\cite{Yu:2016spg}, the differential cross sections for $\gamma n \to K^{\ast 0} \Lambda$ are nearly described by the $t$-channel $K$-trajectory exchange. But there the angular distributions for the $\gamma p \to K^{\ast +} \Lambda$ reaction are only qualitatively described due to the lack of $s$-channel resonance exchanges. In particular, the shapes of the angular distributions near the $K^{\ast +}\Lambda$ threshold exhibited by the CLAS high-precision data, which are suggested to be dominated by the contributions from the $N(2060)5/2^-$ and $N(2000)5/2^+$ resonance exchanges \cite{Wang:2017tpe,Wang:2019mid,Wei:2020},  are missing. 

We mention that in our present work, the contributions from the $t$-channel $K$ exchange are not flexible. Instead, they are totally determined in our previous study of the $\gamma p \to K^{\ast +} \Lambda$ reaction \cite{Wei:2020}, for which much more data on both differential cross sections and spin density matrix elements are available. 
 
In Fig.~\ref{fig:total}, we show the total cross sections of $\gamma n \to K^{\ast 0} \Lambda$ predicted in the present work. For comparison, the total cross sections of $\gamma p \to K^{\ast +} \Lambda$ from Ref.~\cite{Wei:2020} are also shown. The dominant individual contributions are coming from the $t$-channel $K$ exchange, the $s$-channel $N(2060)5/2^-$ exchange, and the $s$-channel $N(2000)5/2^+$ exchange in both reactions, which are plotted with red dotted, green dash-dotted, and blue dashed lines, respectively. The individual contributions from other terms are too small to be clearly seen with the scale used, and thus they are not plotted in Fig.~\ref{fig:total}. It is seen that the contributions from the $t$-channel $K$ exchange dominate the total cross sections of $\gamma n \to K^{\ast 0} \Lambda$, and actually, these contributions are much stronger than those in $\gamma p \to K^{\ast +} \Lambda$ simply because the coupling constant $g_{\gamma K^0 K^{\ast 0}}=-0.631$ has a much larger magnitude than that of $g_{\gamma K^+ K^{\ast +}}=0.413$ as determined by the radiative decays of $K^{\ast 0} \to K^0 \gamma$ and $K^{\ast +} \to K^+ \gamma$. In both reactions, the contributions from the $t$-channel $K$ exchange are close to those from the Born term, indicating negligible contributions from other nonresonant terms. For $\gamma p \to K^{\ast +} \Lambda$, both the $N(2060)5/2^-$ and $N(2000)5/2^+$ exchanges have significant contributions to the cross sections, and the coherent sum of them dominates the total cross sections of this reaction. For $\gamma n \to K^{\ast 0} \Lambda$, the resonance contributions are much weaker but still evident. More specifically, the $s$-channel $N(2060)5/2^-$ exchange provides a little bit weaker contributions in $\gamma n \to K^{\ast 0} \Lambda$ than in $\gamma p \to K^{\ast +} \Lambda$, and the $s$-channel $N(2000)5/2^+$ exchange provides much smaller contributions in $\gamma n \to K^{\ast 0} \Lambda$ than in $\gamma p \to K^{\ast +} \Lambda$.

One sees from Fig.~\ref{fig:total} that our predicated total cross sections of $\gamma n \to K^{\ast 0} \Lambda$ are roughly $1.5$ times larger than those of $\gamma p \to K^{\ast +} \Lambda$. In Ref.~\cite{Wan:2015gsl}, the maximum of the total cross sections was predicated to be around $0.4$ $\mu$b, almost the same as that for $\gamma p \to K^{\ast +} \Lambda$. In Ref.~\cite{Yu:2016spg}, the predicated total cross sections of $\gamma n \to K^{\ast 0} \Lambda$ are roughly $1.1$ times larger than those of $\gamma p \to K^{\ast +} \Lambda$. Experimental data on the total cross sections of $\gamma n \to K^{\ast 0} \Lambda$ are called for to distinguish these theoretical models.

\section{Summary and conclusion}  \label{Sec:summary}

In this work, we have studied the photoproduction reaction $\gamma n \to K^{\ast 0} \Lambda$ based on the same reaction mechanism as in our previous work for the study of the $\gamma p \to K^{\ast +} \Lambda$ reaction. Our purpose is to obtain a unified description of the available differential cross-section data for both the $\gamma p \to K^{\ast +} \Lambda$ and $\gamma n \to K^{\ast 0} \Lambda$ reactions and the data on spin density matrix elements for the $\gamma p \to K^{\ast +} \Lambda$ reaction within the same effective Lagrangian model. It is expected that a combined analysis of all the available data for both these two reactions will put more constraints on the theoretical model and will result in more reliable understanding of the reaction mechanisms of both the $\gamma p \to K^{\ast +} \Lambda$ and $\gamma n \to K^{\ast 0} \Lambda$ reactions. 

The interaction current and the $t$-channel $K^\ast$ exchange vanish automatically for $\gamma n \to K^{\ast 0} \Lambda$ due to the neutral charges of $K^{\ast 0}$ and $n$. Apart from that, the $t$-channel $K$ and $\kappa$ exchanges, the $u$-channel $\Lambda$, $\Sigma$, and $\Sigma^\ast$ exchanges, and the $s$-channel $N$, $N(2060)5/2^-$, and $N(2000)5/2^+$ exchanges are considered in calculating the reaction amplitudes. The hadronic coupling constants, propagators, and the resonance masses and widths are taken from our recent work of Ref.~\cite{Wei:2020} for the study of $\gamma p \to K^{\ast +} \Lambda$. The only adjustable parameters in the present work are the products of the resonance hadronic and electromagnetic coupling constants, $g_{RN \gamma}^{(1)} g_{R\Lambda K^\ast}^{(1)}$ and $g_{RN \gamma}^{(2)} g_{R\Lambda K^\ast}^{(1)}$, which are determined by a fit to the available differential cross-section data for $\gamma n \to K^{\ast 0} \Lambda$.

The available differential cross-section data for $\gamma n \to K^{\ast 0} \Lambda$ have been reproduced quite well. The numerical results show that the contributions from the $t$-channel $K$ exchange dominate the cross sections of the $\gamma n \to K^{\ast 0} \Lambda$ reaction. Unlike Refs.~\cite{Wan:2015gsl,Yu:2016spg} where the cross sections for $\gamma n \to K^{\ast 0} \Lambda$ are almost fully described by the $t$-channel $K$ exchange and all other contributions are negligible, in the present work the contributions from the $N(2060)5/2^-$ and $N(2000)5/2^+$ exchanges turn out to be rather considerable. The total cross sections of $\gamma n \to K^{\ast 0} \Lambda$ are predicated to be $1.5$ times larger than those of $\gamma p \to K^{\ast +} \Lambda$. More experimental data on this reaction are called for to put further constraints on the theoretical models.

\begin{acknowledgments}
The author Y.M.Z. is grateful to Yu Zhang and Ai-Chao Wang for their useful and constructive discussions. This work is partially supported by the National Natural Science Foundation of China under Grants No.~12175240, No.~11475181, and No.~11635009, the Fundamental Research Funds for the Central Universities, the Key Research Program of Frontier Sciences of Chinese Academy of Sciences under Grant No.~Y7292610K1, and the China Postdoctoral Science Foundation under Grant No.~2021M693142.
\end{acknowledgments}


\end{document}